\documentstyle[aps,prl,twocolumn]{revtex}
\title{Excitation and relaxation in atom--cluster collisions}
\author{U. Saalmann}\address{Max--Planck--Institut f{\"u}r Physik
komplexer Systeme, N{\"o}thnitzer Str.\ 38, 01187 Dresden, Germany}
\author{R. Schmidt}\address{Institut f{\"u}r Theoretische
Physik, Technische Universit{\"a}t Dresden, 01062 Dresden, Germany
\\ [5mm]
\begin{minipage}{400pt}\small
Electronic and vibrational degrees of freedom in atom--cluster 
collisions are treated simultaneously and self--consistently 
by combining time--dependent density functional theory with 
classical molecular dynamics. The gradual change of the
excitation mechanisms (electronic and vibrational) as well as the
related relaxation phenomena (phase transitions and fragmentation)
are studied in a common framework as a function of the impact energy
(eV\ldots MeV).
Cluster ``transparency'' characterized by
practically undisturbed atom--cluster penetration is predicted 
to be an important reaction mechanism 
within a particular window of impact energies.
\\ [1mm]
\end{minipage}}

\input{epsf}\epsfclipon
\begin{document}
\maketitle

Collisions with atomic clusters represent a relatively new
branch of collision physics as compared to the well established
fields of ion--atom collisions \cite{br83}
and ion--solid interaction \cite{nama+96}.
The study of cluster collisions is of particular interest and
importance because it offers the possibility to tackle
bridge--building questions (like the continuous transition from
individual excitations in the elementary ion--atom collision
to the macroscopic stopping power in solids) 
as well as fundamental problems (like phase transitions in
finite systems).
It is also an extremely challenging and complicated field as the
comprehensive understanding of these collisions still requires
the development of basically new techniques for both,
large--scale (multi--parametric) experiment and many--body
(quantum--mechanical) theory.
In this request, the present situation resembles very much 
that of nuclear physics at the beginning of the 80's \cite{bo80}.

Experimentally, large progress has been made, meanwhile, in the 
investigation of \emph{adiabatic\/} cluster collisions where the 
reaction mechanism is determined by vibrational excitations only.
Typical examples are the study of the vibrational energy transfer
\cite{bukr94}, the fusion between clusters \cite{casc+93}, 
the formation of endohedral complexes \cite{wehr+91},
and the collision induced dissociation (CID) \cite{jabo+87}.
There is also a lasting interest to study \emph{non--adiabatic\/} 
cluster collisions where electronic transitions occur.
Experiments in this field concern the measurements of 
the charge transfer \cite{brca+88,waco+94,shhv+95},
ionization and electronic excitation \cite{chgu+95},
as well as the \emph{selective\/} observation of vibrational
and electronic excitations \cite{brdu+96}.

Theoretically, adiabatic cluster collisions can be well described by 
quantum molecular dynamics (QMD) or molecular dynamics (MD)
\cite{knsc97,roca+96,icik+96,zhwa+94,robr+95}.
Also, isomeric \cite{jebo+94} and
solid--liquid phase transitions \cite{be94a,kito94} in clusters 
as well as the fission process of clusters \cite{brca+94}
have been studied with MD or QMD where, basically, electronic
excitations are not considered. 
On the other hand, electronic transitions 
in non--adiabatic cluster collisions have been treated with
classical \cite{base95,th95}, 
semi--classical \cite{grgu95}, 
or one--electron quantum mechanical \cite{boha93,gusi97} approaches 
where the atomic structure, and thus the 
vibrational degrees of freedom, are not taken into account.
Recently, a general theory has been developed \cite{sasc96}
which is able to describe simultaneously adiabatic and
non--adiabatic collisions and, in particular, also the still
completely unknown transition regime where both~--- electronic
and vibrational~--- excitations occur.
This so--called non--adiabatic quantum molecular dynamics
(NA--QMD) \cite{sasc96} can also be used to study, for the first
time, phase transitions in and fragmentation of clusters induced
by electron--vibration coupling. 

In this work, different excitation mechanisms (electronic and
vibrational) as well as related relaxation phenomena (phase
transitions and fragmentation) in cluster collisions are studied
in the microscopic framework of NA--QMD.
This theory treats electronic and vibrational degrees of
freedom simultaneously and self--consistently
in atomic many--body systems by combining time--dependent density
functional theory \cite{rugr84} with classical MD.
The key point, in order to derive practicable equations of
motion, is the split of the total electronic density into the
adiabatic and a remaining part, treating afterwards the
exchange--correlation terms with the adiabatic density only
\cite{sasc96}. 
This approximation has been applied and successfully tested
against experimental data, so far, for the interpretation of
fragment correlations in CID \cite{faba+98}
and the calculation of absolute cross sections for charge
transfer \cite{tobe1} in cluster reactions.
Here the universal NA--QMD approach is used to investigate
the gradual change of the excitation and relaxation
mechanisms as a function of the impact energy in a wide
range for two, basically different collision systems,
Na$_9{\!}^+$\,+ Na (with attractive adiabatic forces between
projectile and target)
and Na$_9{\!}^+$\,+ He (with dominating repulsive forces and a
``magic'' initial electronic configuration).

In Fig.~1, the total kinetic energy loss (TKEL) of the collision,
$\Delta E{=}E_{\rm cm}-E_{\rm cm}(t{\to}+\infty)$,
with $E_{\rm cm}$ the impact energy and 
$E_{\rm cm}(t{\to}+\infty)$
the final kinetic energy of the relative motion between
cluster--projectile and atomic target in the center--of--mass
system is shown calculated for a fixed collision geometry (with
impact parameter $b{=}0$) but in a wide range of impact energies
$E_{\rm cm}{=}0.2$\,eV\ldots1\,MeV. 
\begin{figure}[t]
\epsfbox{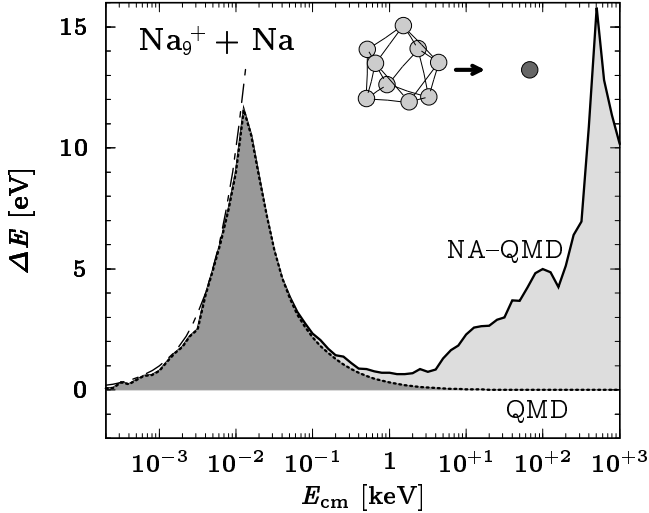}
\caption{Total kinetic energy loss 
of relative motion $\Delta E$  in central Na$_9{\!}^+$\,+ Na 
collisions  for a fixed collision geometry (see insert)
as a function of the center--of--mass impact energy $E_{\rm cm}$
calculated with  NA--QMD (solid line) and QMD (dotted line).
Vibrational and electronic contributions to $\Delta E$ are distinguished
by grey and light--grey shaded areas, respectively.
The dot--dashed line corresponds to $\Delta E=E_{\rm cm}$.}
\end{figure}
For ``real'' scattering events (see below), $\Delta E$ describes
the change of the internal energy of the system generally
connected with excitations in the cases studied here.
In order to decide between electronic and vibrational contributions
to $\Delta E$ 
the NA--QMD results are compared to those obtained from
adiabatic QMD \cite{sasc96} calculations performed with the same
collision geometry.
At impact energies below 
$E_{\rm cm}\stackrel{<}{\scriptstyle\sim}200$\,eV, 
QMD and NA--QMD results are clearly seen to be identical and,
thus, only vibrational excitations occur (adiabatic regime).
Above $E_{\rm cm}\stackrel{>}{\scriptstyle\sim}5$\,keV 
electronic transitions dominate (non--adiabatic regime).
In the intermediate range of $E_{\rm cm}$ both mechanisms
compete (transition regime).
Maximal TKEL, namely $\Delta E{=}E_{\rm
cm}$, is practically realized at all impact
energies below $E_{\rm cm}\stackrel{<}{\scriptstyle\sim}10$\,eV
and is connected with the formation of  relatively long--living
but in general unstable 
intermediate compounds Na$_{10}{\!}^+$ \cite{tobe}.
The ``real'' scattering events with incomplete energy loss
appear above $E_{\rm cm}\stackrel{>}{\scriptstyle\sim}10$\,eV.

In Fig.~2, the time dependence of 
$\Delta E(t){=}E_{\rm cm}{-}E_{\rm cm}(t)$, with $E_{\rm cm}(t)$ the
actual kinetic energy of the relative motion, 
and the displacement of the cluster atoms $d(t)$ are shown for
characteristic impact energies.
The quantity $\Delta E(t)$ gives insight into the collision and
excitation dynamics (forces, interaction time, TKEL).
The displacement, defined as
$d(t){:=}${}$ \left[{\textstyle\sum_{A{=}1}^N} \left[{\bf R}_A(t)
-{\bf R}_A(0)\right]^2\right]^{1/2}\!\!\Big/(NR)$
with ${\bf R}_A(0)$ the equilibrium positions of the $N$ atoms
in the ground state configuration and $R$ the cluster radius, 
characterizes quantitatively
the relaxation, i.\,e., if $d$ is well below one a ``solid''
configuration is described, if $d$ lies in the vicinity of one a
``liquid'' state is realized, and if $d$ goes to infinity
fragmentation occurs.
\begin{figure}
\epsfbox{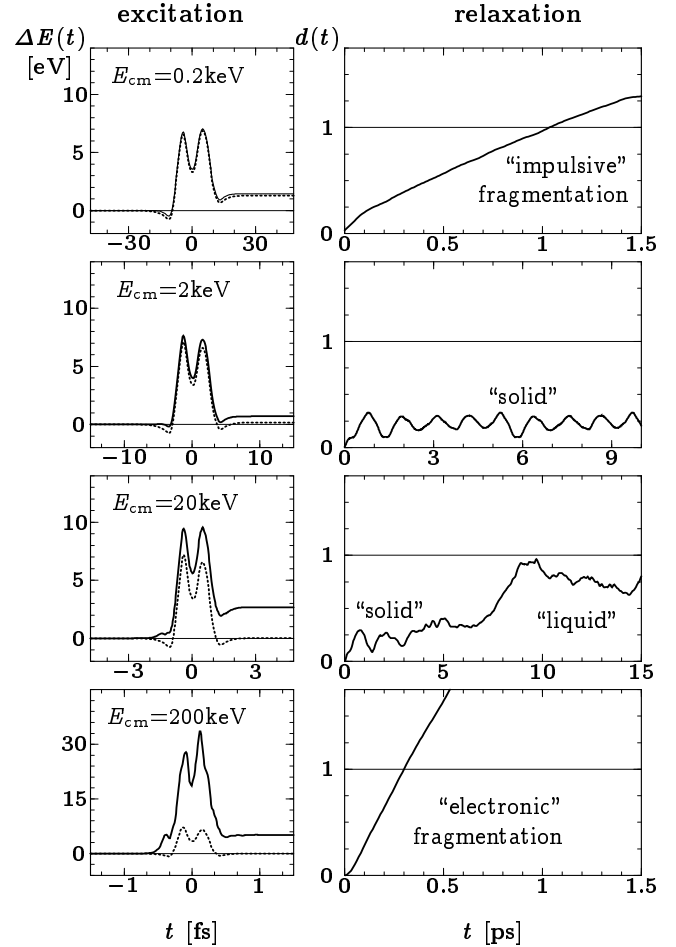}
\caption{Difference of the impact energy and the actual kinetic energy 
of relative motion $\Delta E(t)$ (left column)
and displacement of the cluster atoms $d(t)$ (right) as a function of time $t$ 
for four impact energies $E_{\rm cm}$=0.2, 2, 20, 200\,keV  and the 
same collision geometry as used in Fig.~1, 
calculated with NA--QMD (dark solid lines).
In the left column, the corresponding adiabatic QMD calculations (dotted lines)
are also shown for comparison.
Note the different time scales for the excitation processes (femtoseconds)
and the relaxation phenomena (picoseconds).}
\end{figure}
Excitation and relaxation mechanisms and, in particular, the
related time scales are basically different in all cases.
In the pure adiabatic regime ($E_{\rm cm}${=}200\,eV), the
vibrational excitation of the cluster ($\Delta E{\approx}1.4$\,eV)
is immediately connected with a fragmentation process which, 
in contrast to statistical evaporation,
starts to proceed already during the interaction time of 
$\sim$\,40\,fs (see left part of Fig.~2)
as a consequence of sufficient momentum transfer between
projectile and target atoms.
Such non--statistical fragmentation has been experimentally
verified only very recently \cite{brdu+96}
(called ``impulsive'' fragmentation mechanism).
On the contrary, in the high--energetic non--adiabatic region
$E_{\rm cm}${=}200\,keV 
electronic excitation ($\sim$ 1\,fs) and fragmentation ($\sim$ 300\,fs)
processes are separated by two orders of magnitude.
Obviously, electron--vibration coupling needs time to become
effective and to induce dissociation
(``electronic'' fragmentation).  
At an impact energy of $E_{\rm cm}${=}2\,keV the cluster remains
stable after the collision with a total excitation energy of
$\Delta E{\approx}0.7$\,eV originally stored predominantly in
electronic excitation  (cf.\ adiabatic and non--adiabatic
contributions to $\Delta E$ in the left part of Fig.~2).
Electron--vibration coupling leads, however, to
the excitation of collective surface vibrations 
indicated by the regular oscillating behaviour of $d(t)$ well
developed after about 1\,ps.
At $E_{\rm cm}${=}20\,keV the first step of the relaxation phase
is very similar to that observed at $E_{\rm cm}${=}2\,keV.
But, in this case electron--vibration coupling induces
clearly a phase transition from ``solid'' to ``liquid'' in a
second relaxation step coming about 10\,ps after the
collision. 
Owing to the large TKEL ($\Delta E{\approx}2.7$\,eV) one should
expect statistical evaporation as the final decay channel.

\begin{figure}
\epsfbox{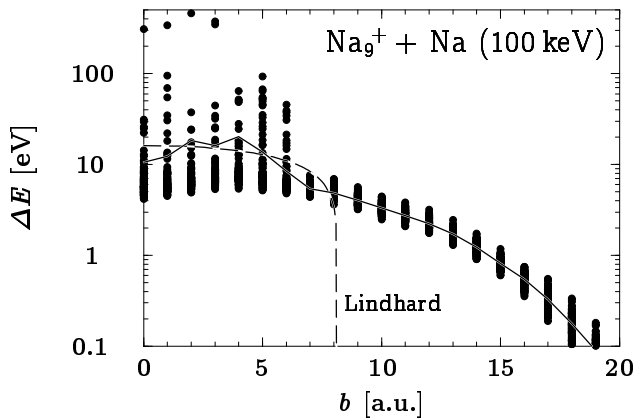}
\caption{Total kinetic energy loss $\Delta E$ 
in Na$_9{\!}^+$\,+ Na collisions ($E_{\rm cm}$=100\,keV)
for randomly chosen cluster orientations as a function 
of the impact parameter $b$
calculated with NA--QMD (dots) 
as well as the resulting mean value (solid line).
The prediction of the Lindhard model \protect\cite{li54} 
is also shown (dashed).}
\end{figure}
The excitation and relaxation mechanisms, summarized in Figs.~1
and 2, appear in a well defined order as a function of the impact
energy due to the deliberately fixed collision geometry.
For a given impact energy and, in particular, for comparison with
experiment the impact parameter 
dependence of measurable quantities averaged over different
orientations of the cluster with respect to the beam axis is of
interest.
\begin{figure}[t]
\epsfbox{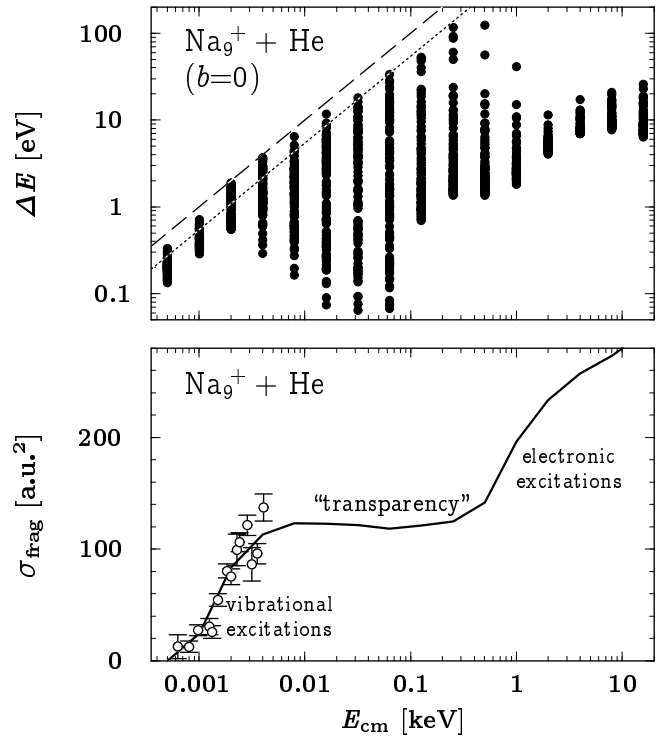}
\caption{Upper part: Total kinetic energy loss $\Delta E$ 
in central Na$_9{\!}^+$\,+ He collisions 
as a function of the center--of--mass impact energy $E_{\rm cm}$
calculated with NA--QMD using randomly chosen orientations (dots).
The lines correspond to a complete energy loss $\Delta E=E_{\rm cm}$
(dashed) and the maximal energy loss for a binary collision
of the He atom with \emph{one} cluster atom (dotted)
\protect\cite{tobe,nota+96a}. 
Lower part: Total fragmentation cross sections as a function 
of $E_{\rm cm}$ calculated with  NA--QMD (solid line).
Available experimental data (circles) are from \protect\cite{nota+96a}.}
\end{figure}
For the TKEL this dependence is shown in Fig.~3 for an impact
energy in the non--adiabatic regime. 
The results are compared to the well--known Lindhard
model \cite{li54}
of the electronic stopping power in solids.
This phenomenological model describes surprisingly well the mean
value of $\Delta E$ for impact parameters smaller than the
cluster radius, $b\stackrel{<}{\scriptstyle\sim}8$\,a.u.
There are, however, two characteristic and measurable
peculiarities of $\Delta E(b)$ in the microscopic calculations.
First, for small impact parameters 
$b\stackrel{<}{\scriptstyle\sim}6$\,a.u., there
are a few events with a very large TKEL.
They result from binary atom collisions between the Na--target
and a cluster atom, leading to very fast fragment monomers.
In a macroscopic language, these direct ``knock--out''
collisions correspond to the nuclear stopping power \cite{nama+96}.
Second, for impact parameters larger than the
cluster radius, $b\stackrel{>}{\scriptstyle\sim}8$\,a.u.,
the calculated (in that case purely electronic) TKEL are not
small as compared to the dissociation energy of the cluster.
Consequently, ``electronic'' fragmentation (cf.\ Fig.~2) may
essentially contribute to the total fragmentation cross section
which can be expected to be several times larger than the
geometrical one.

The most interesting and unexpected phenomenon of the studies
performed so far is the occurrence of cluster ``transparency''
in the closed shell system Na$_9{\!}^+$\,+ He.
This mechanism is characterized by a practically undisturbed
atom--cluster penetration in a certain, small range of 
$E_{\rm cm}$ within the transition regime.
In Fig.~4, the TKEL as a function of $E_{\rm cm}$ calculated
for random orientations with impact parameter $b{=}0$ is shown.
It can be clearly seen that in the window between $E_{\rm
cm}\sim10$\,eV and $\sim$ 100\,eV the variance of $\Delta E$
becomes markedly large and there are a lot of events with an
extremely small energy loss.
Obviously, for specific collision geometries in this range of
$E_{\rm cm}$ the relative velocity between projectile and target
becomes too large to excite vibrational degrees of freedom and
is, on the other hand, still too small to induce electronic
transitions.
As a consequence, the cluster appears to be transparent even in 
central collisions.

One possibility to study the change in the excitation mechanisms
and in particular the occurrence of transparency is to consider
the total fragmentation cross section as a function of 
$E_{\rm cm}$ (lower part of Fig.~4).
The theoretical cross section $\sigma_{\rm frag}$ 
has been calculated by taking into account an initial
temperature of the clusters according to recent experiments
\cite{nota+96a}.  
An excellent agreement between theory and experiment is found in
the adiabatic regime where experimental data are available at present,
i.\,e.\ $E_{\rm cm}\stackrel{<}{\scriptstyle\sim}4$\,eV.
It is just the transparency effect which leads to a
pronounced plateau in $\sigma_{\rm frag}$  between 10\,eV and
100\,eV before electronic excitations induce a further increase
of $\sigma_{\rm frag}(E_{\rm cm})$.

In summary, we have shown how different excitation mechanisms
compete and various relaxation phenomena occur in atom--cluster
collisions.
First relations and basic differences to macroscopic systems
have been discussed.
Cluster transparency has been predicted to be an important
reaction mechanism in Na$_9{\!}^+$\,+ He.
A systematic investigation of different collision systems 
(in particular the size dependence of transparency
\cite{footnote1}, its eventual relation to the channeling effects in
solids \cite{ge74}, the temperature dependence, etc.)
will be the scope of future studies.

We are grateful to O.\ Knospe for helpful discussions
and E.\,E.\,B.\ Campbell for careful reading of the
manu\-script. 
This work was supported by the DFG through
Schwer\-punkt ``Zeitabh\"angige Ph\"anomene \ldots'' and
by the EU through the HCM network CMRX--CT94--0614.


\begin{thebibliography}{10}
\bibitem{footnote1}
Preliminary studies show, e.\,g., that this mechanism is much less pronounced
  in Na$_8{\!}^+$\,+ He \cite{tobe}.
\end{thebibliography}
\end{document}